\documentstyle[11pt,paspconf,epsf]{article}

\begin{document}

\title{ Regularities in galaxy spectra and the nature of the Hubble sequence}
\author{ Laerte Sodr\'e Jr. and H\'ector Cuevas}
\affil{Departamento de Astronomia, Instituto Astron\^omico e Geof\'{\i}sico 
da Universidade de S\~ao Paulo, SP, Brazil}

\begin{abstract}
We have investigated some statistical properties of integrated spectra of
galaxies with principal component analysis.
The projection of the spectra onto the plane defined by the first two principal
components  shows that normal galaxies are in a quasi-linear 
sequence that we call spectral sequence and 
is closely related to the Hubble morphological sequence.
 We verify that the spectral sequence is also an evolutive
sequence, with galaxy spectra evolving from later to earlier spectral types.
Considering the close correspondence between the spectral and 
morphological sequences, we speculate that galaxies may evolve 
morphologically along the Hubble sequence, from Sm/Im to E.
If this is the case, the first galaxies were mainly gas rich irregular
objects, which evolved later along the morphological sequence as long as
mergers and interactions increased their masses and developed their
spheroidal components.
\end{abstract}


\section{Introduction}
We present a study of Kennicutt's (1992a,b) spectrophotometric 
atlas, looking for global regularities in the integrated spectra of galaxies. 
This is done with a standard statistical technique, Principal Component 
Analysis (PCA). Similar analysis, dealing mainly with spectral classification,
are presented by Connolly et al. (1995) and Folkes, Lahav \& Maddox (1996).

Kennicutt's atlas contains 55 integrated spectra. 
We consider here a subsample with 23 normal galaxies. This set 
covers the Hubble 
sequence from E to Im, avoiding objects with any evidence of peculiarity 
(e.g. AGNs, starbursts, mergers). Although small, the sample discussed here
allows us to explore the connection between normal galaxy spectra and
morphology. An analysis of all spectra is presented
elsewhere (Sodr\'e \& Cuevas 1996).

\section{Principal Component Analysis}
Suppose we have a sample of N integrated spectra, all covering the
same rest-frame wavelength range (3750 - 6500 \AA). 
Each spectrum is described by a $M$-dimensional vector
containing the galaxy flux at $M$ uniformly sampled wavelengths
(here $M$=1300).
Let ${\cal S}$ be the $M$-dimensional space spanned by these `spectral' vectors.
An integrated spectrum, then, is a point in the ${\cal S}$-space.
Here we apply PCA to find a suitable plane in ${\cal S}$, defined by the
two first principal components of the data, $y_1$ and $y_2$.
We call it the {\it principal plane}, because it is the plane that contains 
most of the variance in the spectral space. 

 We apply PCA to the covariance
matrix of the data, and with spectra normalized in mean flux, $\sum_\lambda
f_\lambda = 1$. 
The total variance  contained in the principal plane is then 93\%.
We show elsewhere (Sodr\'e \& Cuevas 1996) that our main
results do not depend on data normalization or whether one applies PCA
to the covariance or to the correlation matrix of the data.

\section{The spectral sequence}
Figure 1a shows the projection of the 23 spectra onto their principal plane. 
The projected spectra of
almost all galaxies are arranged along a quasi-linear sequence which we  
call {\it spectral sequence}. 
The few outliers are Sc and Sm/Im galaxies with spectra dominated by nebular
emission lines.
We also show in figure 1a the projections of the mean spectra
binned in 5 morphological groups: E, S0, Sa-Sbc, and Sc-Im.
They preserve, over the spectral sequence, the same ranking of the
Hubble morphological sequence. We conclude that {\it the spectral sequence
correlates strongly with the Hubble morphological sequence} (see also
Sodr\'e \& Cuevas 1994; Connolly et al. 1995; Folkes, Lahav \& Maddox 1996).
Note that the
spectral sequence provides quantitative support for a one-dimensional
description of the general properties of normal galaxies, like the Hubble
morphological sequence. It also indicates that there is not a dividing line 
in global spectral properties between
elliptical and spiral galaxies, and that lenticulars have spectra intermediate
between those two morphological groups. 

The very existence of the spectral sequence, and its correlation with the 
Hubble sequence, indicates that one single parameter
may be responsible for the integrated spectra of normal galaxies- and of
the morphological sequence. For 
instance, Gallagher, Hunter \& Tutukov (1984), 
Sandage (1986) and Ferrini \& Galli (1988),
suggest that several properties of the Hubble sequence can be explained by
variations in the star formation rate of galaxies. 

We have investigated this hypothesis with Bruzual \& Charlot (1995, hereafter
B\&C) revised models of spectral evolution 
of galaxies. We have only considered models
with a single parameter, the characteristic star formation 
time-scale of a galaxy, $\tau$.
In figure 1b we show, for several values of $\tau$, the projection of the
model spectra onto the principal plane defined by the 23 galaxies in our
set of normal galaxies. 
The model spectra overlap partially the spectral sequence for
all but the latest types. For these galaxies, emission lines associated to
ongoing star formation dominate the spectra.

We plot in figure 1c the evolutive track for the $\tau=0$  
model, with galaxy ages running from 0 to 20 Gyr. 
The evolutive track of an instantaneous burst overlaps most of
the spectral sequence, that is, its spectrum evolves from late to 
early-types along the sequence. 
For the constant star formation rate model ( $\tau=\infty$), the evolutive 
track overlaps the $\tau=0$ one, but in this case the oldest spectrum falls 
near the centroid of normal Sc galaxies.  Hence, the
spectral sequence not only characterizes the locus of normal galaxies 
(except for those whose spectra are dominated for nebular lines) but
is almost coincident with their evolutionary tracks.
Our results lead us to conclude that {\it the spectral sequence
is also an evolutive sequence, with normal galaxy spectra evolving from 
late-type irregulars to that of ellipticals.} 
Consequently, we expect that the fraction of galaxies with late-type spectra
should increase with redshift. 

\begin{figure}
\plotfiddle{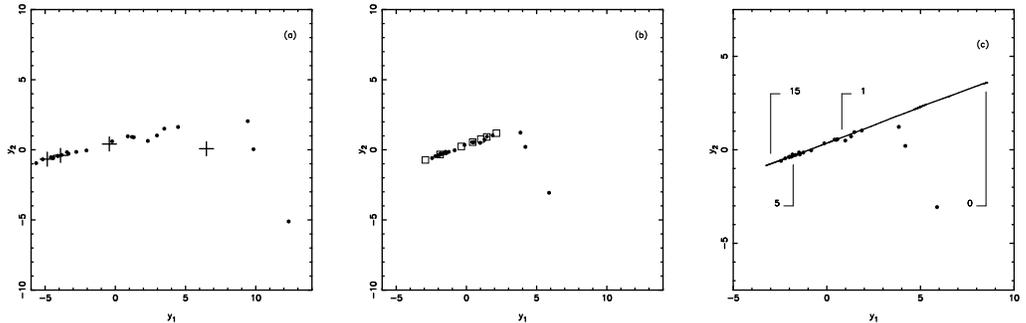}{5cm}{270}{70}{70}{-200}{300}
\caption{Projection of the spectra of normal galaxies onto the
principal plane (dots). a) the projections of the mean spectra of
E, S0, Sa-Sbc, and Sc-Im (from left to right) are represented by crosses.
b) projection of B\&C model spectra for several values of the mean star
formation time scale $\tau$ and age of 15 Gyr 
(open squares): from left to right $\tau=$ 0, 3, 5, 7, 10, 15 and
$\infty$ Gyr. c) projection of B\&C evolutive track for $\tau=0$, with the
age in Gyr indicated next to the track.} \label{fig-1}

\end{figure}

\section{The nature of the Hubble sequence}
We have seen that there is a close correspondence
between the spectral and morphological sequences today, and
that the spectral sequence is also an evolutive sequence.
If we do not live in a special epoch in the history of the Universe, we may
suppose that the spectral and the morphological sequences have been
always related. Then the Hubble sequence would be also an evolutionary
 sequence, 
with galaxies evolving from Im to E! This is a speculation, of course, since
our results refer to spectra and not to morphologies, and then the appearance
of an E galaxy with age of 1 Gyr may be quite different of what is today an
$\sim$Sc galaxy. But is worth noting that some recent works present evidence
that  the morphology of a normal galaxy may evolve from 
late to early types. For instance, Pfenniger, Combes \& Martinet (1994) and
 Pfenniger, Martinet \& Combes (1996) argue that
this evolution may be driven by internal and external factors due to the
likely coupling between dynamics and star formation, as long as galaxies are
able to accrete mass. 
The key point
is that dynamical process that actuate during
a galaxy life- like formation and destruction of bars, mergers, 
close encounters, gas compression and/or stripping, etc.-
tend to  favour an increase of the 
spheroidal component at the expense of the
disk, leading to a univocal sense of 
morphological evolution, from Sm to Sa, and eventually to S0 or E.

This sense of evolution also explains the morphological content of galaxy
clusters, where most of galaxies are E or S0. The simplest
explanation assumes that there is an infalling population of late-type
galaxies (Sodr\'e et al. 1989, Kauffmann 1995) that are transformed in 
E and S0 (as well as in dwarf ellipticals) in the hostile environment
of the clusters. Moore et al.
(1996) have recently proposed that frequent encounters at high speed among
the galaxies in clusters (``galaxy harassment") may be the driver of
morphological transformations in these environments. This process may explain
the Butcher-Oemler effect and the form of the blue objects observed in
high-redshift clusters by the HST.

Hierarchical models also indicate that galaxy morphology may well change 
as consequence of mergers and interactions (Baugh, Cole \& Frenk 1996),
although these models do not have yet enough resolution to address the question
of evolution within spiral types.  

This scenario also
 suggests that the first galaxies should look more like faint gas
rich irregular objects- in agreement with the inhomogeneous galaxy formation
models of Baron \& White (1987)-
 than the presumed bright precursors of today's
elliptical galaxies and bulges of spirals in the scenario devised by
Eggen, Lynden-Bell \& Sandage (1962).

\end{document}